hep-ph/9311012
\documentstyle[12pt]{article}

\newlength{\dinwidth}
\newlength{\dinmargin}
\newcommand{\resection}[1]{\setcounter{equation}{0}\section{#1}}

\begin{document}
\vspace*{7cm}
\begin{center}
  \begin{Large}
  \begin{bf}
Effective action method for computing
%the $1/N$
next to leading corrections of
$O(N)$ models \\
  \end{bf}
  \end{Large}
  \vspace{5mm}

  \begin{large}
D. Dominici\\
  \end{large}
Dipartimento di  Fisica, Univ. di Firenze\\
I.N.F.N. Sezione di Firenze\\
  \vspace{5mm}

  \begin{large}
U. Marini Bettolo Marconi\\
  \end{large}
Dipartimento di  Matematica e Fisica, Univ. di Camerino\\
%I.N.F.N., Sezione di Perugia e I.N.F.M. Sezione di Camerino\\
I.N.F.M. Sezione di Camerino\\
  \vspace{5mm}

\end{center}
  \vspace{2cm}
\begin{center}
DFF 193/9/1993\\\
\end{center}
\vspace{1cm}
\noindent

\newpage
\thispagestyle{empty}
\begin{quotation}
\vspace*{5cm}
\begin{center}
\begin{bf}
ABSTRACT
\end{bf}
\end{center}
\vspace{5mm}
We compute the corrections of
%order $O(N)$ and $O(1)$
next to leading order in the ${1 \over N}$
expansion to the effective
potential of a system described by a Ginzburg-Landau model with $N$
components and quartic interaction, in the case of
spontaneous symmetry breaking. The method
we apply allows to generalize in a simple way the so-called
Self-Consistent Screened Approximation (SCSA).
%Moreover it prooves to be useful in the computation of the $\eta$
%exponent in
%the case of interactions of higher order.

\vspace{5mm}
\noindent
\end{quotation}
\newpage
\setcounter{page}{1}
\def\lq{\left [}
\def\rq{\right ]}
\def\bfi{\bar \phi}
\def\ddk{ \frac{d^Dk}{(2\pi)^D}}
\def\ddq{ \frac{d^Dq}{(2\pi)^D}}
\def\dmu{\partial_{\mu}}
\def\dmus{\partial^{\mu}}
\def\Gh{\hat {\cal G}}
\def\GG{{\cal G}}
\def\pa{\phi_{\alpha}}
\def\DD{{\cal D}}
\def\Tr{{\rm Tr}}
\def\eps{{\epsilon}}
\newcommand{\be}{\begin{equation}}
\newcommand{\ee}{\end{equation}}
\newcommand{\bea}{\begin{eqnarray}}
\newcommand{\eea}{\end{eqnarray}}
\newcommand{\nn}{\nonumber}
\newcommand{\dd}{\displaystyle}

\resection{The CJT formalism}

\par
 Variational principles provide a very useful tool for studying field
theories and a valid alternative to perturbative methods.
The simplest of these principles is perhaps the Hartree
method, which consists in finding the best quadratic approximation
to a given Hamiltonian, by rendering the variational estimate of the
effective potential minimal with respect to some adjustable parameters.
 In order to go beyond the quadratic approximation several methods have been
devised, such as two or three loop approximations post-gaussian methods
etc. \cite{Stevenson} . In the present paper,
we derive and generalize, to the case of a spontaneously broken symmetry,
the so-called Self Consistent Screened Approximation (SCSA),
originally introduced by Bray \cite{Bray}, in the case of an $O(N)$ invariant
model by means of a general method introduced by
Cornwall, Jackiw and Tomboulis (CJT) \cite{CJT} in Quantum Field Theory (see
also \cite{domokos} for previous work).
Our analysis makes also use of the $\frac{1}{N}$
expansion method which has been already
employed in the $O(N)$ theory \cite{Coleman}\cite{altri}\cite{Parisi} .
 We believe that the compactness
of the present derivation provides a simpler route to obtain results
for generic $O(N)$ models.

The basic idea, behind this effective action method for composite
operators, is to introduce, in the generating functional, sources
which couple to the composite operators, to be studied, and then
perform a double Legendre transform.

Let us consider for the sake of generality the generating
functional for the Green functions
\be
Z[J,K]=\int\DD \phi \exp \left [-S (\phi)+J\phi
+{1\over 2} \phi K \phi\right ]
\ee
where $J$ and $K$ are respectively a local and a bilocal
source:
\be
J \phi = \int  d^Dx J(x) \phi (x)
\ee
and
\be
{1\over 2} \phi K \phi ={1\over 2} \int  d^Dx \int d^Dy
\phi (x) K(x,y) \phi (y)
\ee

By considering $W=-\ln Z$ and defining the variables
\be
\phi_c={\delta W\over \delta J}~~~~~~~~~~~~
\GG={{\delta^2 W}\over {\delta J \delta J}}=
2 {{\delta W}\over {\delta K}}-\phi_c\phi_c
\ee
one can eliminate the sources $J$ and $K$ in favor of
the fields $\phi_c$ and $\GG$ and obtain the
generating functional for 2PI (Two Particle Irreducible)
Green functions
\be
\Gamma (\phi_c,\GG)=W-J\phi_c-
{1\over 2} \phi_c K \phi_c -{1\over 2} \GG K
\ee
2PI meaning that the graphs cannot be disconnected
by cutting only two propagator lines.

The stationarity conditions for the functional $\Gamma$ read:
\be
{{\delta \Gamma }\over {\delta \phi_c}}=-J-K\phi_c~~~~
{{\delta \Gamma }\over {\delta \GG}}=-{1\over 2}K
\ee
and physical processes correspond to vanishing sources
$J=0$ and $K=0$.

As shown in \cite {CJT} it is possible to get a formal series for $\Gamma$
\be
\Gamma (\phi_c,\GG)=-S(\phi_c)-{1\over 2} \Tr \ln \GG^{-1}
-{1\over 2}\Tr \DD^{-1}\GG +\Gamma_2 (\phi_c,\GG)
\ee
where
\be
\DD^{-1}(\phi_c;x,y)=\frac {\delta^2 S(\phi)}{\delta \phi (x)\delta \phi (y)}
\vert{\dd\phi_c}
=D^{-1}(x,y)+\frac{\delta^2 S_{int}(\phi)}{\delta \phi (x)\delta \phi (y)}
\vert_{\dd\phi_c}
\ee
$\Gamma_2$ is given by the 2PI vacuum diagrams of a theory
with interactions determined by $S_{int}$ and propagators
$\GG$, where $S_{int}$ is defined by the shifted action
\be
S(\phi_c+\psi)-S(\phi_c)-\psi \frac{\delta S} {\delta \phi_c}=
{1\over 2}\psi \DD^{-1}\psi +S_{int}(\psi,\phi_c)
\ee
Notice that this procedure represents a dressed loop expansion, all the
propagators being fully radiative corrected and can thus
exhibit non perturbative effects even for a small number of
dressed loops. The second of eqs.
(1.6) together with eq. (1.7) represents the Schwinger-Dyson equation for
the propagator $\GG$
\be
\GG^{-1}=\DD^{-1}+\Sigma (\phi_c,\GG)
\ee
where the last term is the self-energy contribution
\be
\Sigma (\phi_c,\GG)=-2\frac{\delta \Gamma_2(\phi_c,\GG)} {\delta \GG}
\ee
As a final remark, frequently one is interested in translationally
invariant solutions $\phi (x)=\phi$ and $\GG (x,y)=
\GG (x-y)$. One therefore defines the so called
effective potential
\be
V(\phi,\GG)=-\Omega_D \Gamma(\phi,\GG)\vert_{trans.~inv.}
\ee
where $\Omega_D$ is a volume factor.

\resection{The effective potential for the $O(N)$ model}

We apply, now, the above construction to
the Ginzburg-Landau $O(N)$ model with quartic interaction, which
is defined by the classical action
\be
S(\phi)=\int d^Dx\lq {1\over 2}\sum_{\alpha}(\partial_\mu \phi_\alpha)^2
+{1\over 2}\mu^2 \sum_{\alpha}\phi_{\alpha}^2
+\frac{\lambda} {4!N}(\sum_{\alpha}\phi_{\alpha}^2)^2\rq
\ee
%where $\phi^2=\sum_{\alpha=1}^{N}\phi_\alpha \phi_\alpha$.

One constructs first the shifted
action and then obtains $\Gamma_2$ in the following way.
One starts by drawing all graphs  at a given order in
$\lambda$, using the vertices determined by the shifted action
(eq.1.9) and propagator $\GG$, and neglecting all
diagrams which are two-particle reducible.

To illustrate the previous rules we consider the
effective action up to $\lambda^2$ order. The list of
graphs contributing to $\Gamma_2$
is given in Fig. 1. As it is well known
in the limit $N\to\infty$ the Hartree approximation,
which corresponds to considering only the first diagram
of Fig. 1, becomes exact. However, if  one
considers the ${1 \over N}$ corrections, one sees that
an infinite set of 2PI diagrams must be kept.

For future applications we are mainly interested in the effective
potential, therefore we consider translationally
invariant solutions $\phi (x)=\phi$ and $\GG (x,y)=
\GG (x-y)$.
Under such hypothesis,
 the following decomposition of the tensor $\GG$ is convenient:
\be
\GG_{\alpha \beta}(x-y)= {\phi_{\alpha}  \phi_{\beta} \over \phi^2 }
\GG_L(x-y)+
\lq \delta_{\alpha\beta}-{\phi_{\alpha}\phi_{\beta}
\over \phi^2 } \rq\GG_T(x-y)
\ee
where $\phi^2=\sum_{\gamma} \phi^2_{\gamma}$.

Therefore we get
\bea
{1\over 2} \Tr \ln \GG^{-1}&=&{1\over 2}(N-1) \Tr \ln \GG_T^{-1}\nn\\
&&+{1\over 2} \Tr \ln \GG_L^{-1}
\eea
\bea
{1\over 2}\Tr \DD^{-1}\GG &=&{1\over 2}\int d^D x\int d^D y
[ (N-1) ( -\Delta +\mu^2+ \frac {\lambda}{6N}\sum_{\alpha}
\phi_{\alpha}^2)\delta (x-y)
\GG_T(x-y)\nn\\
&&+  ( -\Delta +\mu^2+\frac {\lambda}{2N}\sum_{\alpha}
\phi_{\alpha}^2  )\delta (x-y)
\GG_L (x-y) ]
\eea

where we have used the inverse propagator
\bea
\DD^{-1}_{\alpha \beta}(x,y)= \lq\Bigl ( -\Delta +\mu^2+{\lambda \over 6 N}
\sum_\gamma\phi^2_{\gamma}\Bigl)\delta_{\alpha\beta}+{\lambda \over 3 N}
{\phi_\alpha\phi_\beta} \rq \delta(x-y)
\eea
and
\bea
-\Gamma_2(\phi,\GG)&=&\frac {\lambda N}{24}\int d^Dx \GG_T^2(0)\nn\\
&&+\frac {\lambda}{12}\int d^Dx \GG_T(0)\GG_L(0)\nn\\
&&-\frac {\lambda^2 \phi^2}{36N}\int\int d^Dx d^Dy
\GG_T^2(x-y)\GG_L(x-y)\nn\\
&&-\frac {\lambda^2}{144}\int\int d^Dx d^Dy\GG_T^4(x-y)
\eea
%We defined
%\be
%\phi_\alpha= \phi_1\delta_{1\alpha}+\psi_\alpha
%\ee
%with $\langle \psi_\alpha\rangle =0$ and
%\be
%\GG_L=\langle \psi_1\psi_1\rangle  ~~~~~
%\GG_T=\langle \psi_\alpha\psi_\alpha\rangle ~~~(\alpha\neq 1)
%\ee
Eq. (2.6) is exact to order $\lambda^2$ and to all orders in $N$.

A more systematic way of improving upon the Hartree approximation
is to observe an interesting aspect of the effective potential,
i.e. that the
leading corrections in the ${1 \over N}$ expansion
of the effective action $\Gamma$ form two simple classes of
diagrams with respect to all possible graphs determined by  $S_{int}$.

In Fig. 2 we draw  the series for $\Gamma_2$ up to $O(N^0)$ to all
orders in $\lambda$. We have neglected graphs like the one
in Fig. 3, which gives a contribution $O(1/N)$.

The $O(N^0)$ truncated series has a simple structure
and it is therefore possible to resum it completely. In fact, one recognizes
$\Gamma_2$ to be the  sum of
two types of contributions:
the first containing only
two trilinear vertices and an arbitrary number of
quadrilinear vertices, the second
only quadrilinear vertices.
By simple
combinatorics one can write the symmetry factors of
each diagram contained in Fig. 4. The symmetry factor
associated to the diagrams in Fig. 4a and 4b are respectively given
by
\be
2^{2n-3}(n-2)!~~~~~~~~~~~~~~~~~~~~~~~~~~~~~~2^{2n-1}(n-1)!
\ee
Thus we get the following contribution from the amplitude corresponding
to Fig. 4a to $\Gamma_2$ (we assume a constant
field $\phi_\alpha (x)=\pa$)
\be
\Gamma_2^{a(n)}={(-1)^n }\frac {\phi^2 }{N}({\lambda \over 6})^n\int
dx_1 \ldots \int dx_{n-1} \int dx_n \GG_L(x_1-x_n)
\GG_T^2(x_1-x_2)\ldots \GG_T^2(x_{n-1}-x_n)
\ee
and
\be
\Gamma_2^{b(n)}=\frac {(-1)^n } {2n}({\lambda\over 6})^n
\int dx_1 \ldots \int dx_{n-1} \int dx_n \GG_T^2(x_1-x_n)
\GG_T^2(x_1-x_2)\ldots \GG_T^2(x_{n-1}-x_n)
\ee
where $n \geq 2$ from the amplitude corresponding to Fig. 4b.
Under the assumption of translation invariance the series corresponding
to eq. (2.8) turns out to be a geometrical series and thus we can recast
the contribution in the following simple form
\be
{\Gamma_2^a\over \Omega_D}= \frac {\phi^2\lambda}
{6N}\int\frac{d^Dq}{(2\pi)^D}
\Gh_L(q)-\frac{\phi^2\lambda}{6N}\int\frac{d^Dq} {(2\pi)^D}
\frac{\Gh_L(q)}{1+{\dd{\lambda\over 6}}\Pi(q)}
\ee
where one has introduced the so called vacuum polarization
\be
\Pi(q)=\int \frac {d^Dk} {(2\pi)^D}\Gh_T(k)\Gh_T(-k-q)
\ee
and $\Omega_D$ is a volume factor.
In an analogous way one gets for the contribution
 from eq.(2.9)
\be
{\Gamma_2^b\over \Omega_D}= \frac {\lambda} {12}\int\frac {d^Dq} {(2\pi)^D}
\Pi(q)-{1\over 2}\int\frac{d^Dq} {(2\pi)^D}\ln\lq
1+{\dd{\frac{\lambda}{6}}} \Pi(q)\rq
\ee
This last result was already obtained by Bray \cite{Bray}.
Notice that the factor $\lq 1+{\dd{\frac{\lambda}{6}}} \Pi(q)\rq^{-1}$
is reminiscent of the improved propagator for the auxiliary field
\cite{Parisi}. In fact, an alternative way to study $\phi^4$ model
is by the trick of the auxiliary field. The new field $\alpha(x)$
is coupled to $\phi$ via trilinear vertices and has a tree-level
propagator equal to 1 (in given units).
Finally taking into account all the contribution we get the final
expression for the effective potential
\bea
V=-{\Gamma\over\Omega_D} &=&{1\over 2}\mu^2\phi^2+{\lambda\over {24N}}
\phi^4\nn\\
&&+{1\over 2}\int \frac{d^Dk}{(2\pi)^D}\Gh_L(k)
\lq k^2+\mu^2+\frac {\lambda}{2N} \phi^2\rq\nn\\
&&+{N-1\over 2}\int \frac{d^Dk}{(2\pi)^D}\Gh_T(k)
\lq k^2+\mu^2+\frac {\lambda}{6N} \phi^2\rq\nn\\
&&-{1\over 2}\int \frac{d^Dk}{(2\pi)^D}\ln \Gh_L(k)
-\frac{N-1}{ 2}\int \frac{d^Dk}{(2\pi)^D}\ln\Gh_T(k)\nn\\
&&+{{\lambda N} \over 24}
\int\ddq\ddk \Gh_T(k)\Gh_T(k-q)\nn\\
&&+{\lambda \over 12}\int\ddq\ddk \Gh_L(k)\Gh_T(k-q)\nn\\
&&-\frac{\phi^2\lambda} {6N}\int \ddq
\Gh_L(q)+\frac {\phi^2\lambda}{6N}\int\ddq
\frac{\Gh_L(q)} {1+{\dd{\lambda\over 6}}\Pi(q)}\nn\\
&& -\frac {\lambda} {12}\int\ddq
\Pi(q)+{1\over 2}\int\ddq\ln\lq
1+{\dd{\lambda\over 6}} \Pi(q)\rq
\eea
We notice that the previous expression must be taken up
to $O(N^0)$ in the large $N$ limit.

To conclude in order to determine the equilibrium value of $\Gamma$
one has to solve explicitly the stationarity conditions of
CJT (see eqs. (1.6) with vanishing external sources). By
differentiating with respect to the fields
$\phi$, $\GG_T$ and $\GG_L$ one gets (when the vacuum expectation
value of $\phi_\alpha$ is different from zero)
\be
\mu^2+{\lambda\over {6N}}\phi^2+{\lambda\over {6N}}
\lq \GG_L(0)+(N-1)\GG_T(0)\rq +{\lambda\over {3N}}
\int\ddq\frac{\Gh_L(q)}{1+{\dd{\frac{\lambda}{6}}}  \Pi(q)}=0
\ee
\be
\Gh_L^{-1}(p)-\hat{\cal{ D_T}}^{-1}(p)-{\lambda\over 6}\GG_T(0)-
{{\lambda\phi^2}\over{3N}}\frac{1}
{1+{\dd{\lambda\over 6}}\Pi(p)}=0
\ee
\bea
&\Gh_T^{-1}(p)-\hat{\cal{ D_T}}^{-1}(p)&-
{\lambda\over {6}}\GG_T(0)-
{\lambda\over {6(N-1)}}
\lq \GG_L(0)-\GG_T(0)\rq\nn\\
&&-{{\lambda}\over{3(N-1)}}\int\ddq\frac{\Gh_T(p-q)}
{1+{\dd{\lambda\over 6}} \Pi(q))}\nn\\
&&+{{\lambda^2\phi^2}\over{9N(N-1)}}\int\ddq\frac{\Gh_L(q)\Gh_T(p-q)}
{\lq 1+{\dd{\lambda\over 6}} \Pi(q)\rq^2}=0
\eea
where
\be
\GG_{T,L}(0)=\int\ddq \Gh_{T,L}(q)
\ee
 In order to solve the coupled set of eqs. (2.14) (2.16)
one proceeds by expanding about
the zeroth order, $O(N^0)$,  Hartree solution.  Since the effective
action contains $\GG_L$ in terms which are $O(N^0)$ we write
\bea
\phi^2&=&N(\chi^2_0+\frac{1}{N}\chi_1^2)\nn\\
\GG_T&=&\GG_T^{(0)}+\frac{1}{N}\GG_T^{(1)}\nn\\
\GG_L&=&\GG_L^{(0)}
\eea
Substituting in eqs.(2.14-16) we find
\bea
&&\mu^2+{\dd{\lambda\over {6}}}\chi^2_0+{\dd{\lambda\over 6}}
\GG_T^{(0)}(0)=0\nn\\
&&\chi_1^2+\GG_L(0)+\GG_T^{(1)}(0)-\GG_T^{(0)}(0)+
2\int\ddq\frac{\Gh_L(q)}{1+{\dd{\frac{\lambda}{6}}}\Pi(q)}=0
\eea
and
\bea
(\Gh_L)^{-1}(p)&=&p^2+\mu^2+
{\dd{\lambda\over {6}}}\chi^2_0+{\dd{\lambda\over 6}}
\GG_T^{(0)}(0)\nn\\
&&+{\dd{\lambda\over {3}}}\frac {\chi^2_0}
{1+{\dd{\lambda\over 6}}\Pi(p)}\nn\\
(\Gh_T^{(0)})^{-1}(p)&=&p^2+\mu^2+
{\dd{\lambda\over {6}}}\chi^2_0+{\dd{\lambda\over 6}}
\GG_T^{(0)}(0)
\nn\\
\Gh_T^{(1)}(p)/\lq\Gh_T^{(0)}(p)\rq^2&=&-{\dd{\lambda\over {6}}}
\lq\chi_1^2+\GG_L(0)+\GG_T^{(1)}(0)-
\GG_T^{(0)}(0)\rq\nn\\
&&-{{\lambda}\over{3}}\int\ddq\frac{\Gh_T^{(0)}(p-q)}
{1+{\dd{\lambda\over 6}} \Pi(q))}\nn\\
&&+{{\lambda^2\chi^2_0}\over{9}}\int\ddq\frac{\Gh_L(q)\Gh_T^{(0)}(p-q)}
{\lq 1+{\dd{\lambda\over 6}} \Pi(q)\rq^2}
\eea

In the absence of magnetic field, i.e. at two phase coexistence
in the thermodynamic language, the
above equations reduce to:
\bea
&&(\GG_L)^{-1}=p^2+{\dd{\lambda\over {3}}}\frac{\chi^2_0}
{1+{\dd{\lambda\over 6}}\Pi(p)}\nn\\
&&(\GG_T^{(0)})^{-1}=p^2
\eea
a result already known \cite{Amit}
and
\bea
\Gh_T^{(1)}(p)/\lq\Gh_T^{(0)}(p)\rq^2&=&{\dd{\lambda\over {3}}}
\int\ddq\frac{\Gh_L(q)}{1+{\dd{\frac{\lambda}{6}}}\Pi(q)}
\nn\\
&&-{{\lambda}\over{3}}\int\ddq\frac{\Gh_T^{(0)}(p-q)}
{1+{\dd{\lambda\over 6}} \Pi(q))}\nn\\
&&+{{\lambda^2\chi^2_0}\over{9}}\int\ddq\frac{\Gh_L (q)\Gh_T^{(0)}(p-q)}
{\lq 1+{\dd{\lambda\over 6}} \Pi(q)\rq^2}
\eea
To summarize we have derived by means of the CJT formalism
the next to leading
order corrections to the effective potential for the $\phi^4$ theory.
The method is easily generalizable to more complex
interactions of the type $(\phi^2)^n$ with $O(N)$ symmetry. The case
$(\phi^2)^3$, relevant for tricritical points, is currently under study.

 It is a pleasure to acknowledge helpful discussions with Amos Maritan.
\par

\par

\newpage
\begin{center}
  \begin{Large}
  \begin{bf}
  Figure Captions
  \end{bf}
  \end{Large}
  \vspace{5mm}
\end{center}
\begin{description}
\item [Fig. 1] Graphs contributing to $\Gamma_2$
               up to $\lambda^2$ order.
\item [Fig. 2] Series of graphs $\Gamma_2$ up to $O(N^0)$ to all
               orders in $\lambda$.
\item [Fig. 3] Example of  neglected graph,
               which gives a contribution $O(1/N)$.
\item [Fig. 4] General graphs of $O(N^0)$ contributing to
               $\Gamma_2$

\end{description}
\end{document}